\documentclass[12pt,preprint]{aastex} 

\lefthead{Escala} 
\righthead{A law for star formation in galaxies.}

\def\msun{M_{\odot}}

\begin{document}

\title{A law for star formation in galaxies.} 
 
\author{Andr\'es Escala} 
\affil{Departamento de Astronom\'{\i}a, Universidad de Chile, Casilla 36-D, Santiago, Chile.} 
\affil{Kavli Institute for Particle Astrophysics and Cosmology, Stanford University Physics Department / SLAC, 2575 Sand Hill Rd. MS 29, Menlo Park, CA 94025, USA.} 
 
\begin{abstract} 
We study the galactic-scale triggering of star formation. We 
 find that the largest mass-scale not stabilized by rotation, a well 
 defined quantity in a rotating system and with clear 
 dynamical meaning, strongly correlates with the star formation rate 
 in a wide range of galaxies. We find that this relation can be 
 understood in terms of self-regulation towards marginal Toomre stability and the amount 
 of turbulence allowed to sustain the system in this self-regulated quasi-stationary state. We
 test such an interpretation by computing the predicted star formation rates for a galactic interstellar medium characterized by lognormal 
 probability distribution function and find good agreement with the observed relation.
 
\end{abstract}


\section{Introduction}

Observations  of normal spiral galaxies by Schmidt (1959)   first 
suggested  that their star formation rates (SFRs) scale with their global properties. This relation was extended to other 
 galaxies with higher SFRs, such as the nuclear  
regions of spiral galaxies  and Ultra Luminous InfraRed Galaxies 
(ULIRGs) by Kennicutt (1998). These  observations have 
lead to an empirical law for star formation, often called the Kennicutt-Schmidt 
(KS) Law: 
\begin{equation} 
 \rm \dot{\Sigma}_{star} \propto \Sigma_{gas}^{1.4} \,\,\, , 
\end{equation}  
where $\rm  \Sigma_{gas}$ and $\rm \dot{\Sigma}_{star}$ are the gas 
surface density and SFR per unit area, respectively.  
 
Since star formation is a local 
process that happens on subparsec scales, the correlation with galactic scale ($>$ 1 kpc)  quantities, such as averaged $\rm  \Sigma_{gas}$,  suggests the 
 existence of a physical connection between galactic and 
subparsec scales. Motivated by the observed  KS law,  several  
authors have tried to find a scenario in which a global/large scale property of   
galaxies could trigger and/or regulate  star formation  (Quirk 
1972; Wyse 1986; Kennicutt 1989;   
Elmegreen 2002; Li et al. 2005). The KS law may also be explained in
terms of processes that are primarily local, within star-forming clouds (e.g. Krumholtz et al. 2009 and references therein). 
 
 From the study of   gravitational 
instabilities in disks, the `turbulent' Toomre  parameter $\rm Q_{turb} \equiv v^{rms}_{turb} 
\kappa / \pi G \Sigma_{gas}$ (or related parameters such as the star formation threshold $\rm 
 \Sigma_{crit}$; Kennicutt 1989) 
 arises as a natural candidate for a key triggering 
 parameter. However, the average $\rm Q_{turb}$ in a galaxy is  
observed to be close to 1,  in galaxies ranging from local spirals
(Martin \& Kennicutt 2001) to starbursts such as ULIRGs (Downes \& Solomon 1998). 
Since observed values of $\rm Q_{turb}$ (or $\rm 
 \Sigma_{gas}/\Sigma_{crit}$) range over at most a factor  of a few, the observed range in SFRs per unit area of seven
orders of magnitude is difficult to explain solely in terms of 
this threshold. Therefore, further investigation is  required to
determine what controls  the SFRs on galactic scales.

An important point, not generally addressed, is that, a disk  at the
condition of marginal Toomre stability can have a range of
possible self-regulated  states. For example, the  nuclear disk in a 
starburst with Q $\sim$ 1 is much more turbulent than
the disk of a normal spiral galaxy with Q $\sim$ 1. The goal of this
work is to 
study which galactic property triggers this more turbulent  behavior
 for some  galaxies and why their SFRs  can be orders of magnitude
higher than in more `quiescent' spiral galaxies.

This work is organized as follows. We first review   gravitational instability analysis  in order to
introduce the largest scale not stabilized by rotation,  followed  in
\S 2 by a discussion of the correlation found between this largest  scale and
the SFR. Section 3 presents a physical 
interpretation of the correlation found, in terms of self-regulation
due to feedback
processes. In \S 4  we test such an interpretation by comparing  predictions
against the observed SFRs. Finally in \S 5, we summarize the  results of this work.

\section{The Maximum Scale not Stabilized by Rotation}

In order to introduce the largest  unstable length scale in  galactic
disks, we  first  review  some standard results from gravitational 
instability analysis (Toomre 1964; Goldreich \& Lynden Bell 1965). For 
one 
of  the simplest cases of a differentially 
rotating thin sheet or disk, linear stability analysis yields the  dispersion relation for small perturbations (Binney 
\& Tremaine 2008)  $\omega^2 = \kappa^2 - 2\pi G \Sigma_{\rm gas}
|k| + k^2 C_{\rm s}^{2} $, where $C_{\rm  s} = \sqrt{dP/d\Sigma}$ is the sound speed, 
$\Sigma$ is the surface density, and $\kappa$ is the epicyclic frequency 
 given by $\kappa^2 (R) = R \frac{d\Omega^2}{dR} \, 
+ \, 4 \Omega^2$ ($\Omega$ being the angular frequency). The system becomes unstable when $\omega^2 < 0$, which is equivalent 
to the condition   Q $<$ 1, where Q is the Toomre parameter and is  
defined as  $\rm  C_{\rm s} \kappa / \pi G \Sigma_{\rm gas}$ . In such a case there 
is a range of unstable length scales limited on small scales by 
thermal pressure (at the Jeans length $\lambda_{\rm Jeans} = C_{\rm
  s}^2 / G\Sigma_{\rm gas} $) and on large scales by rotation (at the critical
length set by rotation, $\lambda_{\rm rot}= 4\pi^2 G \Sigma_{\rm gas} /
\kappa^2$).  All intermediate length scales are unstable, and the most
rapidly growing mode has a wavelength $2 \, \lambda_{\rm Jeans}$
(Binney \& Tremaine 2008; Escala \& Larson 2008).

The maximum unstable length scale in a disk, $\lambda_{\rm rot}$, is a robust quantity because it depends only on the surface 
density and epicyclic frequency of the disk and not on smaller scale 
physics. Such a length scale has an associated  characteristic mass, defined 
as equal to  $ \Sigma_{gas} (\lambda_{\rm rot}/2)^{2}$, which can be expressed as: 
\begin{equation} 
\rm M_{\rm rot} = \frac{4\pi^4G^2\Sigma_{\rm gas}^3}{\kappa^4} \, . 
\label{mrot} 
\end{equation} 

On the other hand, due to the  complex structure and dynamics 
of the real interstellar medium (ISM) in galaxies, which cannot be described by a simple equation of state, 
there is not a 
well-defined  Jeans length at intermediate scales. Therefore, there is no real lower 
limit on the sizes of the self-gravitating structures that can form 
until the  thermal Jeans scale is reached in molecular cloud cores 
(Escala \& Larson 2008). 

The combination of the observed correlations of SFRs with galactic properties, and the fact that  this largest scale not stabilized by rotation is the only well defined galactic scale in the gravitational instability
problem, is our motivation for exploring a possible  link between SFR and this characteristic galactic scale.
 

\subsection{The Law}

In order to test the  existence of a link between the largest scale not stabilized by rotation and the SFR in galaxies, we will check  whether   the mass-scale 
defined by rotation (Eq. 2) correlates with the SFR. For a rotationally supported system, the average of this mass-scale  can be expressed 
in terms of quantities such as the gas mass and gas fraction (Escala \& Larson 2008): 
\begin{equation} 
\rm M_{rot}  = 3 \, \times \, 10^7\,\msun \,\, \frac{ M_{\rm gas}}{10^9 \,\msun} \left(\frac{\eta}{0.2}\right)^2, 
\label{eq2} 
\end{equation} 
where $\rm M_{\rm gas}$ is the  gas mass in the disk and  $\rm 
\eta = M_{\rm gas}/M_{\rm dyn}$ is the ratio of the gas mass to the 
 total enclosed dynamical mass within the gas radius (this varies from 
 the disk radius for spiral galaxies to the radius of the nuclear starburst  
 disk/ring in ULIRGs). This expression (Eq. \ref{eq2}) has the advantage  that it reduces 
the scatter due to error propagation compared to the original formulation (Eq. \ref{mrot}). 
 
\begin{figure}
   \centering
   \includegraphics[width=8.9cm]{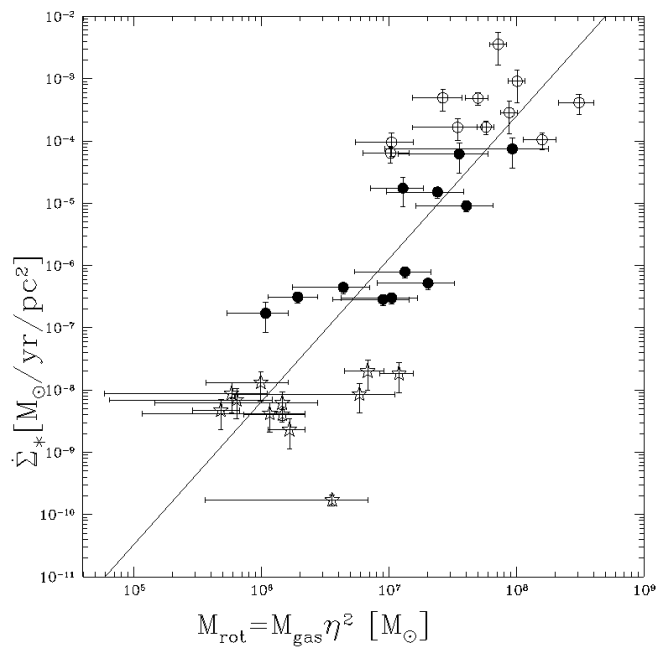}
   \caption{ The  star formation
     rate  plotted against the  critical mass-scale defined by rotation
     $\rm M_{rot}$,   estimated from Eq. \ref{eq2}
     using measured quantities in those galaxies. The open circles show data 
     for nuclear starburst  disks,
     the star symbols show for normal spirals galaxies and the  filled
     circles show for
     nuclear gas in spirals. The solid line corresponds
     to $\rm \rm \dot{\Sigma}_{\star}  \propto M_{rot}^{2.3}$.
}
   \label{SFR}
 \end{figure}

In Figure \ref{SFR} we plot   the maximum mass scale not stabilized by 
rotation estimated from Eq. \ref{eq2} against the measured SFR (per unit area)  
in those galaxies.
 We plot  normal spirals as star symbols, the nuclear gas in normal 
spirals  as filled circles, and ULIRGs as open circles. For the
computation of $\rm M_{rot}$ we have considered only molecular gas
masses and molecular gas fractions. This is because  molecular gas
should be intimately related to the SFR because it is this gas which
eventually forms the stars in Giant Molecular Clouds (GMCs). Error bars displayed in
Figure \ref{SFR} show uncertainties estimated by  error 
propagation from the  uncertainties found in the literature.

 The information for each  data point plotted in
Figure \ref{SFR} is listed in Table 1, together with a list of
references to the works in the literature from which the values were taken. For the spiral galaxies,  $\rm  
 \dot{\Sigma}_{\star}$ is estimated from the $\rm  H_{\alpha}$ luminosity, the gas masses are estimated by the CO 
luminosity, and the dynamical masses are estimated using the 
method listed in   Table 1. For the nuclear gas in normal 
spirals, the gas masses  are estimated  using the CO luminosity, the
dynamical masses from rotation curves  and $\rm  
 \dot{\Sigma}_{\star}$ is estimated using  method listed in 
 Table 1. Finally, for ULIRGs  $\rm  \rm \dot{\Sigma}_{\star}$ is estimated from the far-infrared luminosity, gas
 masses are estimated from the  CO luminosity and dynamical masses are estimated using rotation curves.

Figure \ref{SFR} shows a clear correlation between 
$\rm M_{rot}$ and the SFR per unit area, which supports the idea that   this threshold  
mass has a relevant role  in the triggering of star formation on galactic scales. The solid black line in Figure 
\ref{SFR} shows  a least-squares fit to the points in the figure  and corresponds to 
 a star formation law of $\rm  \rm \dot{\Sigma}_{\star}  \propto M_{rot}^{2.3}$,  with a
 scatter of 0.21 dex. This relation  has a level of scatter comparable
 to the typical scatter found for the KS law. Moreover, this  is the only 
correlation of the SFR with a galactic quantity with
a clear dynamical meaning in terms of stability analysis, and
therefore with a clear role in the star formation problem. 

In summary, the  correlation between the SFR and the maximum 
unstable mass defined by rotation is indeed observed in galaxies, over a 
range that spans almost eight orders of magnitude in SFR per unit area.

\begin{table}
\small

\begin{center}
\caption{Galaxy Physical Parameters}

\resizebox{12cm}{!}{
\makebox[\textwidth]{

\begin{tabular}{lllllc}
\hline
Galaxy & $\rm \dot{\Sigma}_\star$ & $\rm M_{H_{2}}$ & $\rm M_{dyn}$ & $\rm \eta=\frac{M_{H_{2}}}{M_{dyn}}$ & References\\
& $\rm M_\odot yr^{-1} kpc^{-2}$ & $\rm 10^9M_\odot$ & $\rm 10^{10}M_\odot$ &  &\\
\hline
\\
NGC6946         & 0.0132 & 3.3 & 19$^a$ & 0.02 & (1), (3)\\
NGC4419         & 0.00404 & 4 & 9$^b$ & 0.044 & (2), (4), (5)\\
NGC4535         & 0.00017 & 6.918 & 26.3$^c$ & 0.0263 & (2), (4), (6) \\
NGC5033         & 0.00229 & 6.708 & 37.1$^b$ & 0.018 &  (1), (7), (8)\\
NGC4254         & 0.01995 & 13.427 & 51.6$^b$ & 0.019 & (1), (4), (9)\\
NGC7331         & 0.004677 & 4.94 & 21.5$^a$ & 0.023 & (1), (8), (10)\\
NGC4303         & 0.018197 &  10 & 25$^a$ & 0.04  & (1), (4), (11)\\
NGC4647         & 0.0069 & 2.63 & 14.6$^c$ & 0.018 & (1), (4), (6)\\
NGC4654         & 0.0087 & 3.236 & 20.88$^c$ & 0.0155 &  (1), (4), (6)\\
NGC4321         & 0.00851  & 14.791 & 64.3$^c$ & 0.023  & (1), (4), (6)\\
NGC4501         & 0.006166 &  9.77 & 69.3$^c$ & 0.0141 & (1), (4), (6)\\
NGC4689         & 0.00417  &  3.09 & 13.8$^c$ & 0.0224   & (1), (4), (6)\\

\hline
NGC253          & 17.378$^d$ & 0.14 & 0.04 & 0.35 & (1), (12)\\
NGC1614         & 61.6595$^d$ & 2 & 1.3 & 0.154 & (1), (13)\\
NGC470          & 0.7892$^e$ & 0.2 & 0.067 & 0.299 & (14)\\
NGC4102 $^*$    & 9.06539$^e$ & 0.8 & 0.3 & 0.26 & (14)\\
NGC4102 $^{**}$ & 3.00654$^e$ & 1.4 & 1.4 & 0.1 & (14)\\
NGC3504 $^*$    & 1.50835$^e$ & 0.8 & 0.4 & 0.2 & (14)\\
NGC3504 $^{**}$         & 2.8349$^e$ & 1.2 & 1.2 & 0.1 & (14)\\
NGC4536         & 0.5175$^e$ & 1.2 & 0.8 & 0.15 & (14)\\
NGC3351         & 0.4421$^f$ & 0.2 & 0.117 & 0.171 & (14)\\
NGC3627         & 0.1698$^d$ & 2 & 6 & 0.03 & (1), (15), (16) \\
NGC6240         & 74.131$^d$ & 3.1 & 1.55 & 0.2 & (1), (17) \\
NGC5005         & 0.3107$^g$ & 1.03 & 2 & 0.05 & (2), (18) \\

\hline
IRAS00057       & 95.1 & 1.4 & 1.4 & 0.1 & (19)\\
IRAS02483       & 63.6 & 1.9 & 2.235 & 0.085 & (19)\\
IRAS10565       & 104.1 & 4 & 1.739 & 0.23 & (19)\\
Mrk231          & 492 & 1.8 & 1.286 & 0.14 & (19)\\
Arp193          & 487.1 & 2.6 & 1.625 & 0.16 & (19)\\
Arp220 disk     & 166.5 & 3 & 1.875 & 0.16 & (19)\\
Arp220 total    & 285.4 & 5.2 & 3.467 & 0.15 & (19)\\
Arp220 west     & 3555.1 & 0.6 & 0.15 & 0.4 & (19)\\
Arp220 east     & 905.7 & 1.1 & 0.314 & 0.35 & (19)\\
IRAS17208       & 413.4 & 6.1 & 2.346 & 0.26 & (19)\\
IRAS23365       & 164.8 & 3.8 & 3.455 & 0.11 & (19), (20)\\

\hline
  * @300pc
  ** @1300pc \\
  a) HI rot. curves
  b) CO rot. curves
  c) $\rm M_{dyn}-L_H$ relation
  d) $\rm L_{FIR}$
  e) $\rm L_{RC}$
  f) $\rm L_{Br \gamma} $
  g) $\rm  L_{H\alpha}$
  \\

\end{tabular}

}}

\label{table1}
\end{center}

(1) Kennicutt (1998); (2) Komugi et al. (2005); (3) Crosthwaite \& Turner (2007); (4) Young et al. (1996); (5) Kenney et al. (1989); (6) Decarli et al. (2007); (7) P\'erez-Torres and Alberdi (2007); (8) Helfer et al. (2002); (9) Sofue et al. (2003); (10) Thilker et al. (2006); (11) Schinnerer et al. (2002); (12) Mauersberger et al. (1996); (13) Alonso-Herrero (2001); (14) Jogee, Scoville and Kenney (2005); (15) Reuter et al. (1996); (16) Warren et al. (2010); (17) Engel et al. (2010); (18) Sakamoto et al. (2000); (19) Downes and Solomon (1998); (20) Murphy et al. (1996)

\end{table}

\section{A Physical Interpretation in Terms of Self-Regulation}

It is well established that  
galactic disks are  observed to be globally in equilibrium, for galaxies
ranging from  local spirals (Martin \& Kennicutt 2001) to starbursts such as ULIRGs (Downes \& Solomon 1998), with measured 
 Toomre Q parameters  close to  1, in the case of  a `turbulent' version of the Toomre parameter: 
$\rm  Q_{turb} = \rm v^{rms}_{turb}  \kappa / \pi G \Sigma$, where $\rm 
v^{rms}_{turb}$ is the observed velocity dispersion in the gas. This state close to 
stability ($\rm Q_{turb} \sim$ 1) was first suggested by Goldreich \& Lynden Bell (1965) to 
be due to a self-regulation  feedback loop; if $\rm Q_{turb} >>$ 1, in  
the absence of heating  driven by  instabilities the disk 
will cool rapidly and the system will 
eventually become unstable, while if  $\rm Q_{turb} <<$ 1 then
instabilities and star 
formation feedback will be so efficient that enough turbulence will be
produced to  `heat' the disk towards   
$\rm Q_{turb} \sim$ 1.

While most galactic disks are close to marginal Toomre stability, some,
such as nuclear disks in starbursts, can be much  
 more turbulent than the disks of normal spiral galaxies. The 
 reason  is that although  disks are all at  $\rm Q_{turb} \sim 1$,
 they can have 
 self-regulated states with different levels of turbulence 
 $(\rm v^{rms}_{turb})$. This can be  easily visualized by considering
 that  the condition 
 $\rm Q_{turb} \sim 1$ implies a mass scale of    $\rm M_{rot}
  \sim  \frac{4 \pi}{ G \kappa} [v^{rms}_{turb}]^3$  using Eq 2. Since
  the epicyclic frequency  $\rm \kappa$ varies only between $\Omega$ and $2\Omega$ for centrally concentrated disks, in a disk supported vertically by turbulence ($\rm
  \Omega^{-1} = R/v_{circ}  \sim H/v^{rms}_{turb}$) it is straightforward to derive  a mass scale of
\begin{equation}
\rm M_{rot} \sim \frac{8\pi}{3G} H
  [v^{rms}_{turb}]^2
\label{mrotvturb}
\end{equation} 
for a median epyciclic frequency  of $\rm \kappa = 3/2 \, \Omega
$. Since   the disk scale-height H is a monotonically increasing  function of
  $\rm v^{rms}_{turb}$ for a
  disk supported vertically by turbulence, the velocity dispersion of turbulent motions follows   $\rm v^{rms}_{turb}   \propto  M_{\rm rot}^\eta$
 with $\rm \eta > 0 $, for a disk with   $\rm Q_{turb} \sim 1$.  Therefore, since some disks have a larger mass-scale not stabilized by 
rotation $\rm M_{\rm rot}$, their large-scale conditions  require
that  
 feedback processes  produce more turbulence  in order to achieve
 $\rm Q_{turb} \sim 1$. 

The existence of disks with self-regulated states of different  
$\rm v^{rms}_{turb}$ is particularly important because it is believed that 
turbulence has a role in
enhancing and  controlling star formation (Elmegreen 
2002; Krumholz \& McKee 2005; Wada \& Norman 2007). In its simplest  
form (proposed by 
Elmegreen 2002), the SFR depends on the probability distribution 
function (PDF) of the gas density produced by galactic turbulence, 
which appears to be lognormal in simulations of turbulent molecular 
clouds and the ISM (Ballesteros-Paredes \& Mac Low 
2002; Padoan \& Nordlund 2002; Kravtsov 2003; Mac Low et al. 2005; Wada \& Norman 2007; 
Wang \& Abel 2009). Moreover,  numerous numerical
studies  support the fact  that the 
dispersion of the lognormal PDF is  determined by the 
rms Mach number  of the turbulent motions (V\'azquez-Semadeni 1994; 
Padoan et al. 1997;  Padoan \& Nordlund 2002; Federrath et 
al. 2008, 2010). Therefore, it is expected that the SFR
in galaxies scales with the velocity dispersion of turbulent
motions. From this we can infer, at least qualitatively, that a
galaxy with a higher $\rm M_{rot}$ should have a  higher star formation
activity. 


In summary, as the largest mass-scale not stabilized by
rotation $\rm M_{rot}$ increases for a given disk, its 
 self-regulated  state  has an increasingly turbulent ISM 
($\rm v^{rms}_{turb}  \propto  M_{\rm rot}^\eta$ with $\rm \eta > 0 $,
 for $\rm Q_{turb} \sim 1$). A  self-regulated  state with higher $\rm 
v^{rms}_{turb}$ (which 
itself controls and  enhances  star formation) implies   higher star 
formation activity, and therefore  the  
correlation between the critical mass-scale defined by rotation  and the SFR
is expected. 

\section{Galactic SFR for an ISM characterized by a Lognormal PDF}

In order to quantify the arguments given above, in this section we
will compute the predicted SFR for an ISM dynamically controlled by
turbulence   in
which the rms velocity dispersion of turbulent motions  $\rm
v^{rms}_{turb}$ is determined by $\rm M_{rot}$. We will follow an
analysis analogous to Wada \&
Norman (2007), starting with  the assumption that the density PDF of the multiphase ISM in a
galactic disk  can be represented by a single lognormal function:
\begin{equation}
\rm f(\rho)d\rho = \frac{1}{\sqrt{2\pi}\sigma}
exp\left[-\frac{ln(\rho/\rho_{0})}{2\sigma^{2}}\right] dln\rho \,\,,
\label{PDF}
\end{equation}
where  $\rm \rho_{0}$ is the characteristic density scale and $\rm
\sigma$ is the dispersion of the lognormal PDF. Although there is evidence for  deviations from a lognormal 
function in the tails of the density PDF (Scalo et al. 1998; Federrath et al.  2010), for simplicity we neglect any higher order 
correction.  

If the star formation  occurs only in regions whose  density is higher than a critical
 value ($\rm \rho > \rho_{c}$), the SFR per unit volume on a global scale is given by:
\begin{equation}
\rm \dot{\rho}_{\star} =\epsilon_{c} (G\rho_{c})^{1/2} f_{c} \left\langle \rho
    \right\rangle_{V}  \,\,,
\label{wada}
\end{equation}
where $\rm \epsilon_{c}$ is the efficency of star formation, $\rm
\delta_{c} =  \rho_{c}/\rho_{0}$ is the critical density contrast for
star formation, $\rm \left\langle \rho  \right\rangle_{V} = \rho_{0}
e^{\sigma^{2}/2}$ is the volume-average density and $\rm f_{c} = 0.5
\left[ 1-Erf\left( \frac{ln\delta_{c} - \sigma^{2}}{\sqrt{2}\sigma}
\right)  \right]$ is the mass fraction of gas whose density is higher
  than $\rm  \rho_{c}$ (for a derivation see \S 3 of Wada \& Norman
  2007). Numerous numerical studies have claimed that, in addition to the
  approximately lognormal form of the PDF, the dispersion $\rm
\sigma$  is determined by the rms Mach number $\rm \mathcal{M}_{rms} =
v^{rms}_{turb}/C_{S}$, where $\rm C_{S} $ is the
sound speed (V\'azquez-Semadeni 1994; Padoan et al. 1997; Padoan \& Nordlund 2002; Federrath et 
al. 2008, 2010). When a log-normal PDF is assumed, the $\rm \sigma-\mathcal{M}_{rms}$ relation can be expressed as (Padoan et al. 1997): 
\begin{equation}
\rm \sigma^{2}= ln (1 + b^{2} \, \mathcal{M}^{2}_{rms}) \, ,
\label{shock}
\end{equation}
where b is a parameter that varies from 0.3 to 1 and depends, for example, on the compressive to solenoidal modes of the turbulence 
forcing (Federrath et al. 2010).  Since the forcing mode (and
therefore b) may vary across  regions of the ISM (Federrath et
al. 2010), and since the sound speed certainly does 
vary  across the ISM, it is convenient for our analysis to define  a
volume-averaged, b-weighted, sound speed $\rm \widetilde{C}_{S} = \left\langle  \frac{C_{S}}{b} \right\rangle_{V}$ .

Using  equations \ref{mrotvturb}, \ref{wada}, and \ref{shock}, we can write the SFR per
unit area ($\rm \dot{\Sigma}_{\star} = \dot{\rho}_{\star} \, H$) as
\begin{displaymath}
\rm \dot{\Sigma}_{\star} = 1.72 \times 10^{-5} \frac{M_{\sun}}{ pc^{2}
  yr} \epsilon_{c}\delta_{c}^{1/2}   \left(\frac{\rho_{0}}{ M_{\sun}
    pc^{-3}}\right)^{3/2}
  \left[ 1-Erf\left(
  z(\delta_{c}, M_{rot} \, H^{-1} \widetilde{C}_{S}^{-2}) \right)
  \right] \times 
\end{displaymath}

\begin{equation}
\rm \times \left[ 1.95 \frac{H}{kpc} + \left(\frac{Km \, s^{-1}}{
   \widetilde{C}_{S}}\right)^{2}  \frac{M_{rot}}{10^6 M_{\sun}}\right]  \, , 
\label{escala}
\end{equation}
and 
\begin{equation}
\rm   z(\delta_{c}, M_{rot} \, H^{-1} \widetilde{C}_{S}^{-2}) =  \frac{ln \delta_{c} - 2 ln\left(1 +
  \frac{3G}{8\pi} \frac{M_{rot}}{   H\widetilde{C}_{S}^{2}} \right)}{2\left(ln\left(1 +
  \frac{3G}{8\pi} \frac{M_{rot}}{ H\widetilde{C}_{S}^{2}} \right)\right)^{1/2}} \, .
\label{z}
\end{equation}

Figure 2 shows the  comparison of  the predicted correlation between
$\rm \dot{\Sigma}_{\star} \, and \, M_{rot}$ from Eq \ref{escala}
against the data listed in Table 1, for several 
values for the model parameters $\rm \rho_{0}, \epsilon_{c},
\delta_{c}, \widetilde{C}_{S}$. The thick solid lines in
Figures 2a, b, c and d  represent the predicted $\rm \dot{\Sigma}_{\star} $   for model parameters  $\rm \rho_{0}=1 M_{\sun}
    pc^{-3},
\epsilon_{c}=0.01, \delta_{c}=10^{3},  \, and \, \widetilde{C}_{S}$ equivalent to a temperature of 250 K. The
other four curves in Figure 2a show variations in the predicted SFRs for  $\rm \rho_{0}=10^{-2}, 10^{-1}, 10^{1}  \, and \,  10^{2} \, M_{\sun}
    pc^{-3}$. The  four curves in Figure 2b show SFRs for  $\rm
    \epsilon_{c}=1, 10^{-1}, 10^{-3}$   and   $10^{-4}$. In Figure
    2c the curves show SFRs for  $\rm \delta_{c}=10, 10^{2}, 10^{4}
    \, and \,  10^{5}$. Finally, the  curves in Figure 2d show SFRs
    for  $\rm \widetilde{C}_{S}$ equivalent to temperatures of T=25, 75, 750 and 2500 K.
  
From Figure 2 it can be concluded that with a single set of
parameters (thick solid lines) Eq \ref{escala} is able to successfully reproduce
the whole correlation, in contrast with the analogous work of
Wada \& Norman (2007), which did not reproduce the K-S law with a
single set of parameters and relied on a variable star formation
efficiency $\rm \epsilon_{c}$ (between normal and starburst galaxies) in
order to reproduce the observed data.

   \begin{figure}
   \centering
   \includegraphics[width=8.9cm]{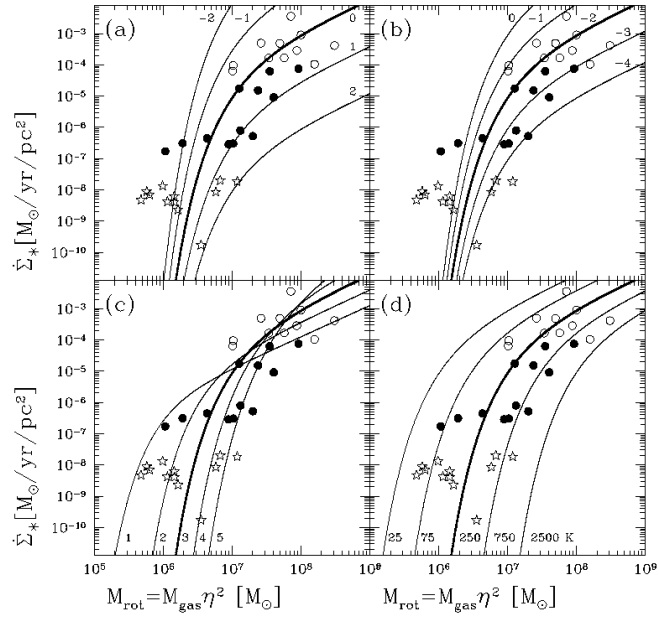}
    \caption{Predicted correlation between $\rm \dot{\Sigma}_{\star} \,
     and \, M_{rot}$ from Eq \ref{escala} against the data listed in 
Table 1, for several 
values for the model parameters $\rm \rho_{0}, \epsilon_{c},
\delta_{c}, \left\langle C_{S}\right\rangle$. The thick solid lines in
all figures show the model with  $\rm \rho_{0}=1 M_{\sun}
    pc^{-3},
\epsilon_{c}=0.01, \delta_{c}=10^{3}  \, and \, \widetilde{C}_{S}$ equivalent to a temperature of 250 K. a) The other lines
 show the variations in SFRs (per unit area) for  $\rm log(\rho_{0}/M_{\sun}
    pc^{-3})= -2, 
-1, 0, 1  \, and \,2 
$. b) The   lines  show SFRs for  $\rm
log\epsilon_{c}= 0, -1, -2, -3  \, and \,  -4$. c) Curves
show SFRs for  $\rm log\delta_{c}= 1, 2, 3, 4  \, and \,
5$. d) Curves show SFRs for  $\rm
\widetilde{C}_{S}$ equivalent to temperatures
  of T=25, 75, 250, 750 and 2500 K.
}
   \label{SFRteory}
   \end{figure}

\section{SUMMARY}

In this Letter we have studied the  role of  the 
 largest mass-scale not stabilized by rotation  in 
galactic disks  in triggering  star formation 
activity in galaxies.

We find that a relation between the largest mass-scale  not
stabilized by rotation and the SFR is 
 observed in galaxies ranging from ULIRGs to normal spirals. This relation  has a  level of scatter   
comparable to  the Kennicutt-Schmidt Law, and is the only known correlation of the global  SFR with a quantity with clear dynamical 
meaning in terms of stability analysis.

We give a physical interpretation for the existence of such a
correlation in terms of self-regulation. In a given disk, as the largest 
 mass-scale not stabilized by rotation increases, its self-regulated 
quasi-stationary state  has  an increasingly  turbulent ISM 
($\rm v^{rms}_{turb}  \propto  M_{\rm rot}^\eta$ with $\rm \eta > 0 $,
 for $\rm Q_{turb} \sim 1$).  Therefore, the role of the critical 
mass-scale in a disk
is to define the  amount of turbulence  allowed to be in
quasi-stationary equilibrium. Since a self-regulated  state with higher $\rm
v^{rms}_{turb}$ enhances  a higher star
formation activity, we expect the existence of a 
correlation between the mass-scale for global stability and the SFR.

We check the validity of this self-regulation  scenario by computing
the predicted SFR  for an ISM dynamically controlled by
turbulence  in which the rms velocity dispersion of turbulent motions  $\rm
v^{rms}_{turb}$ is determined by $\rm M_{rot}$. We find good agreement between the predicted and observed SFRs.

I would like to thank Richard Larson  for 
valuable comments on an early version of the draft, Catherine Vlahakis for
proofreading this manuscript and the referee, Brant Robertson, for a constructive report. I'm indebted to Fernando Becerra for performing the error analysis and graphical display. I also acknowledge partial support from the Center of Excellence in Astrophysics 
and Associated Technologies (PFB 06), FONDECYT Iniciacion Grant
11090216 and from the Comite Mixto ESO-Chile.



 
 

\end{document}